\documentclass[a4paper,preprint,groupedaddress,showpacs,nofootinbib]{revtex4}

\usepackage{amssymb,amsmath,graphicx}

\begin{document}

\title[Lambda Perturbations of Keplerian Orbits]%
{Lambda Perturbations of Keplerian Orbits}

\author{Yurii V. Dumin}

\affiliation{
Sternberg Astronomical Institute (GAISh), Lomonosov Moscow State University,\\
Universitetskii prosp.\ 13, Moscow, 119992 Russia}

\affiliation{
Space Research Institute (IKI), Russian Academy of Sciences,\\
Profsoyuznaya str.\ 84/32, Moscow, 117997 Russia}

\email[]{dumin@yahoo.com, dumin@sai.msu.ru}

\begin{abstract}
To estimate influence of the ``dark energy'' on the Keplerian orbits,
we solve the general relativistic equations of motion of a test
particle in the field of a point-like mass embedded in the cosmological
background formed by the Lambda-term with realistic cosmological
Robertson--Walker asymptotics at infinity.
It is found that under certain relations between three crucial parameters
of the problem---the initial radius of the orbit, Schwarzschild and
de~Sitter radii---the specific secular perturbation caused by
the Lambda-term becomes significant, \textit{i.e.}\ can reach the rate of
the standard Hubble flow.
This fact is interesting both by itself and may have important consequences
for the long-term dynamics of planets and stellar binary systems.
\end{abstract}

%
% 04.25.Nx Post-Newtonian approximation; perturbation theory;
%          related approximations
% 95.10.Ce Celestial mechanics (including n-body problems)
% 04.80.Cc Experimental tests of gravitational theories
% 95.36.+x Dark energy

\maketitle

\section*{1. Introduction}

The question if the planetary orbits and other small-scale celestial systems
are subject to the cosmological influences (in particular, if they feel
the universal Hubble expansion) was put forward by McVittie as early
as 1933~\cite{mvi33}; and this problem attracted attention of a number
of other researchers during the few subsequent decades~%
\cite{ein45,mis73,and95,car98,dom01,dum03,che03,bal06,ior06,kag06,jet06};
a quite comprehensive review of these works was given by Bonnor~\cite{bon00}.

The most frequent conclusion of such studies was that the effect of
cosmological expansion at the planetary scales should be very small
or absent at all. However, the particular estimates given by different
authors substantially disagree with each other.
Moreover, the most of these estimates (excluding the recent ones)
are not applicable to the case when the cosmological background is
formed by the ``dark energy'' (\textit{i.e.}, the $\Lambda$-term in
Einstein equations), because it is distributed perfectly uniform
and insensitive to the local gravitational perturbations.

For example, the most well-known argument against the local Hubble
expansion is the so-called Einstein--Straus theorem~\cite{ein45}:
Let us consider a uniform background cosmological matter distribution
and, next, cut out a spherical cavity and concentrate all its mass
in the central point. Then, a solution of the General Relativity
equations will be given by the purely static Schwarzschild
metric inside the cavity, and it will transform to the time-dependent
Friedmann--Robertson--Walker metric outside the cavity.
In other words, there is no Hubble expansion in the local
empty neighborhood of the point-like massive body, but
such an expansion appears in the regions of space filled with
the cosmological background matter.
Unfortunately, despite an apparent generality of this result,
it is evidently inapplicable to the $\Lambda$-dominated
cosmology, because it is meaningless to consider an empty cavity
in the vacuum energy distribution.

Similarly, it can be shown that such arguments against the local
Hubble expansion as the ``virial criterion'' of gravitational
binding and Einstein--Infeld--Hoffmann surface integral
method~\cite{and95} also do not work when the cosmological background
is formed by the perfectly-uniform $\Lambda$-term.

\section*{2. Theoretical analysis}

\subsection*{2.1. Space--time metric}

From our point of view, the most straightforward and self-consistent
way to estimate how much can the dark energy affect the planetary
dynamics is just to solve the two-body problem in the purely
$\Lambda$-background. In the simplest case of a test particle of
the infinitely small mass moving around the point-like mass~$M$, this
can be done using the well-known solution of General Relativity
equations obtained long time ago by Kottler~\cite{kot18}
(in the modern literature, it is often called
Schwarzschild--de~Sitter solution):
\begin{eqnarray}
& {\rm d} s^2 =
& -\, {\Bigl( 1 - \frac{2 G M}{c^2 r'}
  - \frac{\Lambda {r'}^2}{3} \Bigr)}\, c^2 {\rm d}{t'}^2
\nonumber
\\
&& + \, {\Bigl( 1 - \frac{2 G M}{c^2 r'}
   - \frac{\Lambda {r'}^2}{3} \Bigr)}^{\!\! -1} \!\! {\rm d}{r'}^2 \!
   + {r'}^2 ( {\rm d}{\theta}^2 \!\!\!
   + {\sin}^2{\theta} \, {\rm d}{\varphi}^2 ) \, ;
\label{eq:Kottler_metric}
\end{eqnarray}
for a more general review, see also~\cite{kra80}.
Here, $G$~is the gravitational constant, $c$~is the speed of
light, and primes designate the original Kottler's ``static''
coordinates.

Since metric~(\ref{eq:Kottler_metric}) was derived well before
the birth of the modern cosmology, it suffers from a lack of
the adequate cosmological asymptotics at infinity; namely,
it does not reproduce the standard Hubble flow. Unfortunately,
this fact was ignored in a large number of recent studies.\footnote{
Therefore, such works analyzed only the ``conservative'' effects
caused by the $ \Lambda $-term; while the cosmological influences,
as such, were ignored \textit{a priori}.}
To resolve the above problem, it is necessary to perform
a transformation to the commonly-used cosmological Robertson--Walker
coordinates (represented below by the variables without primes):
\begin{subequations}
\begin{eqnarray}
&& r' = \,
  a_0 \, {\exp} \Bigl( \frac{\displaystyle c t}%
  {\displaystyle r_{\Lambda}} \Bigr) \, r \, ,
\label{eq:r_coord}
\\[1.0ex]
&& t' = \,
  t - \frac{1}{2} \, \frac{\displaystyle r_{\Lambda}}{\displaystyle c} \:\:
  {\ln} \Bigl[ 1 - \frac{\displaystyle a_0^2}{\displaystyle r_{\Lambda}^2} \:
  {\exp} \Bigl( \frac{\displaystyle 2 c t}{\displaystyle r_{\Lambda}} \Bigr)
  \: r^2 \Bigr] \, ,
\label{eq:t_coord}
\end{eqnarray}
\end{subequations}
as outlined in our earlier work~\cite{dum07}.
As a result, the metric will take the form:
\begin{eqnarray}
& {\rm d} s^2 =
& \, g_{tt} \, c^2 {\rm d}{t}^2
  + \, 2 \, g_{tr} \, c \, {\rm d}{t} \, {\rm d}{r}
  + \, g_{rr} \, {\rm d}{r}^2
\nonumber
\\
&& + \, g_{\theta \theta} \, {\rm d}{\theta}^2
   + \, g_{\varphi \varphi} \, {\rm d}{\varphi}^2 \, ,
\label{eq:metric_cosm_coord}
\end{eqnarray}
where
\begin{subequations}
\begin{eqnarray}
&& g_{tt} = \frac{\displaystyle - {\Bigl( 1 - \frac{r_g}{r'} -
  \frac{{r'}^2}{r_{\Lambda}^2}
\Bigr)}^{\!\! 2} + {\Bigl( 1 - \frac{{r'}^2}{r_{\Lambda}^2} \Bigr)}^{\!\! 2}
\frac{{r'}^2}{r_{\Lambda}^2}}%
  {\displaystyle {\Bigl( 1 - \frac{r_g}{r'} - \frac{{r'}^2}{r_{\Lambda}^2}
\Bigr)} {\Bigl( 1 - \frac{{r'}^2}{r_{\Lambda}^2} \Bigr)}^{\!\! 2} } \:\: ,
\label{eq:g_tt}
\\
&& g_{tr} =
  \frac{\displaystyle {\Bigl( 1 - \frac{{r'}^2}{r_{\Lambda}^2} \Bigr)}^{\!\! 2}
- {\Bigl( 1 - \frac{r_g}{r'} - \frac{{r'}^2}{r_{\Lambda}^2} \Bigr)}^{\!\! 2}}%
  {\displaystyle {\Bigl( 1 - \frac{r_g}{r'} - \frac{{r'}^2}{r_{\Lambda}^2}
\Bigr)} {\Bigl( 1 - \frac{{r'}^2}{r_{\Lambda}^2} \Bigr)}^{\!\! 2}}
  \; \frac{r'}{r_{\Lambda}} \, \frac{r'}{r} \:\: ,
\label{eq:g_tr}
\\
&& g_{rr} =
  \frac{\displaystyle {\Bigl( 1 - \frac{{r'}^2}{r_{\Lambda}^2} \Bigr)}^{\!\! 2}
- {\Bigl( 1 - \frac{r_g}{r'} - \frac{{r'}^2}{r_{\Lambda}^2} \Bigr)}^{\!\! 2}
\frac{{r'}^2}{r_{\Lambda}^2} }%
  {\displaystyle {\Bigl( 1 - \frac{r_g}{r'} - \frac{{r'}^2}{r_{\Lambda}^2}
\Bigr)} {\Bigl( 1 - \frac{{r'}^2}{r_{\Lambda}^2} \Bigr)}^{\!\! 2}}
  \; \frac{{r'}^2}{r^2} \:\: ,
\label{eq:g_rr}
\\[1ex]
&& g_{\theta \theta} = \, g_{\varphi \varphi} / {\sin}^2 {\theta} = {r'}^2 \! .
\label{eq:g_angl}
\end{eqnarray}
\end{subequations}
In the above formulas,
$ r_g \! = 2 G M / c^2 $~is Schwarzschild radius,
$ r_{\Lambda} = \sqrt{ 3 / \Lambda } $~is de~Sitter radius, and
$a_0$~is the scale factor of the Universe.

Taking $ a_0 \! = 1 $ at $ t \! = 0 $ and keeping only
the lowest-order terms of $ \, r_g $ and $ 1 / r_{\Lambda} $,
metric~(\ref{eq:g_tt})--(\ref{eq:g_angl})
can be rewritten in a more compact form:
\begin{subequations}
\begin{eqnarray}
&& g_{tt} \approx
  - \Bigl[ 1 - \frac{2 G M}{c^2 r}
  \Bigl( 1 - \frac{c \sqrt{\Lambda} \, t}{\sqrt{3}} \Bigr) \Bigr] \, ,
\label{eq:g_tt_approx}
\\
&& g_{tr} \approx
  \frac{4 \, G M \sqrt{\Lambda}}{\sqrt{3} \, c^2} \:\: ,
\label{eq:g_tr_approx}
\\
&& g_{rr} \approx
  \Bigl[ 1 + \frac{2 G M}{c^2 r}
  \Bigl( 1 - \frac{c \sqrt{\Lambda} \, t}{\sqrt{3}} \Bigr) \Bigr]
  \Bigl( 1 + \frac{2 c \sqrt{\Lambda} \, t}{\sqrt{3}} \Bigr) \, ,
\label{eq:g_rr_approx}
\\
&& g_{\theta \theta} =
g_{\varphi \varphi} / {\sin}^2 {\theta} \approx
  \, r^2 \Bigl( 1 + \frac{2 c \sqrt{\Lambda} \, t}{\sqrt{3}} \Bigr) \, .
\label{eq:g_angl_approx}
\end{eqnarray}
\end{subequations}

\subsection*{2.2. Equations of motion}

Motion of a test particle in
metric~(\ref{eq:g_tt_approx})--(\ref{eq:g_angl_approx})
is described by the geodesic equations which can be derived by
the standard way:
\begin{subequations}
\begin{eqnarray}
&& 2 \: \Big[ 1 - \frac{r_g}{r} \Big( 1 - \frac{t}{r_{\Lambda}} \Big) \Big] \:
\ddot{t} \: - \: 4 \, \frac{r_g}{r_{\Lambda}} \, \ddot{r} \:
+ \: \frac{r_g}{r_{\Lambda}} \, \frac{1}{r} \: {\dot{t}}^{\, 2} \:
\nonumber \\[-0.5ex]
&& \qquad
+ \: 2 \, \frac{r_g}{r^2}
  \Big( 1 - \frac{t}{r_{\Lambda}} \Big) \, \dot{t} \, \dot{r} \:
+ \: \frac{1}{r_{\Lambda}} \, \Big( 2 + \frac{r_g}{r} \Big) \: {\dot{r}}^{\, 2}
\nonumber \\[-0.5ex]
&& \qquad
+ \: 2 \, \frac{r^2}{r_{\Lambda}} \, \Big( \, {\dot{\theta}}^{\, 2}
  + \, {\sin}^2{\theta} \, {\dot{\varphi}}^{\, 2} \Big) \:
= \: 0 \; ,
\label{eq_motion_gen_t}
\\
&& 4 \, \frac{r_g}{r_{\Lambda}} \, \ddot{t} \:
+ \: 2 \, \Big[ 1 + 2 \, \frac{t}{r_{\Lambda}}
  + \frac{r_g}{r} \Big( 1 + \frac{t}{r_{\Lambda}} \Big) \Big] \, \ddot{r} \:
\nonumber \\[-0.5ex]
&& \qquad
+ \: \frac{r_g}{r^2} \Big( 1 - \frac{t}{r_{\Lambda}} \Big) \,
{\dot{t}}^{\, 2} \!
+ \frac{2}{r_{\Lambda}} \Big( 2 + \frac{r_g}{r} \Big) \, \dot{t} \, \dot{r}
- \frac{r_g}{r^2} \Big( 1 + \frac{t}{r_{\Lambda}} \Big) \, {\dot{r}}^{\, 2} \:
\nonumber \\[-0.5ex]
&& \qquad
- \: 2 \, r \Big( 1 + 2 \, \frac{t}{r_{\Lambda}} \Big)
  \Big( \, {\dot{\theta}}^{\, 2} + \, {\sin}^2{\theta} \,
  {\dot{\varphi}}^{\, 2} \Big) \:
= \: 0 \; ,
\label{eq_motion_gen_r}
\\
&& r \Big( 1 + 2 \, \frac{t}{r_{\Lambda}} \Big) \, \ddot{\theta} \:
+ \: 2 \, \frac{r}{r_{\Lambda}} \, \dot{t} \, \dot{\theta} \:
+ \: 2 \, \Big( 1 + 2 \, \frac{t}{r_{\Lambda}} \Big) \, \dot{r} \,
\dot{\theta} \:
\nonumber \\[-0.5ex]
&& \qquad
- \: r \, \Big( 1 + 2 \, \frac{t}{r_{\Lambda}} \Big)
  \sin{\theta} \, \cos{\theta} \, {\dot{\varphi}}^{\, 2}
= \: 0 \; ,
\label{eq_motion_gen_theta}
\\
&& r \Big( 1 + 2 \, \frac{t}{r_{\Lambda}} \Big) \sin{\theta} \, \ddot{\varphi}
+ 2 \, \frac{r}{r_{\Lambda}} \sin{\theta} \: \dot{t} \dot{\varphi}
+ 2 \, \Big( 1 + 2 \, \frac{t}{r_{\Lambda}} \Big)
  \sin{\theta} \, \dot{r} \dot{\varphi} \:
\nonumber \\[-0.5ex]
&& \qquad
+ \: 2 \, r \Big( 1 + 2 \, \frac{t}{r_{\Lambda}} \Big)
  \cos{\theta} \: \dot{\theta} \dot{\varphi} \,
= \: 0 \; ,
\label{eq_motion_gen_phi}
\end{eqnarray}
\end{subequations}
where, for conciseness, we put $ c \equiv 1 $, and dot denotes
a derivative with respect to the proper time~$ \tau $ of the moving
particle.

Finally, if the coordinate system is oriented so that the particle
moves in the equatorial plane, $ \theta = \pi / 2 = {\rm const} $,
then equations~(\ref{eq_motion_gen_t})--(\ref{eq_motion_gen_phi})
are reduced to the following set:
\begin{subequations}
\begin{eqnarray}
&& 2 \: \Big[ 1 - \frac{r_g}{r} \Big( 1 - \frac{t}{r_{\Lambda}} \Big) \Big] \:
\ddot{t} \: - \: 4 \, \frac{r_g}{r_{\Lambda}} \, \ddot{r} \:
+ \: \frac{r_g}{r_{\Lambda}} \, \frac{1}{r} \: {\dot{t}}^{\, 2} \:
\nonumber \\[-0.5ex]
&& \qquad
+ \: 2 \, \frac{r_g}{r^2}
  \Big( 1 - \frac{t}{r_{\Lambda}} \Big) \, \dot{t} \, \dot{r} \:
+ \: \frac{1}{r_{\Lambda}} \, \Big( 2 + \frac{r_g}{r} \Big) \: {\dot{r}}^{\, 2}
\nonumber \\[-0.5ex]
&& \qquad
+ \: 2 \, \frac{r^2}{r_{\Lambda}} \, {\dot{\varphi}}^{\, 2}
= \: 0 \, ,
\label{eq_motion_1}
\\
&& 4 \, \frac{r_g}{r_{\Lambda}} \, \ddot{t} \:
+ \: 2 \, \Big[ 1 + 2 \, \frac{t}{r_{\Lambda}}
  + \frac{r_g}{r} \Big( 1 + \frac{t}{r_{\Lambda}} \Big) \Big] \, \ddot{r} \:
+ \: \frac{r_g}{r^2} \Big( 1 - \frac{t}{r_{\Lambda}} \Big) \,
  {\dot{t}}^{\, 2} \!
\nonumber \\[-0.5ex]
&& \qquad
+ \: \frac{2}{r_{\Lambda}}
  \Big( 2 + \frac{r_g}{r} \Big) \, \dot{t} \, \dot{r} \:
- \: \frac{r_g}{r^2}
  \Big( 1 + \frac{t}{r_{\Lambda}} \Big) \, {\dot{r}}^{\, 2} \!
\nonumber \\[-0.5ex]
&& \qquad
- \, 2 \, r
  \Big( 1 + 2 \, \frac{t}{r_{\Lambda}} \Big) \, {\dot{\varphi}}^{\, 2}
= \: 0 \, ,
\label{eq_motion_2}
\\
&& r \Big( 1 + 2 \, \frac{t}{r_{\Lambda}} \Big) \, \ddot{\varphi} \:
+ \: 2 \, \frac{r}{r_{\Lambda}} \: \dot{t} \dot{\varphi} \:
+ \: 2 \, \Big( 1 + 2 \, \frac{t}{r_{\Lambda}} \Big) \, \dot{r}
  \dot{\varphi} \:
= \: 0 \, .
\label{eq_motion_3}
\end{eqnarray}
\end{subequations}

It should be mentioned that, in the case of a realistic planetary
system, the problem under consideration involves three characteristic
scales, which differ from each other by many orders of magnitude:
Schwarzschild radius~$ r_g $ (\textit{e.g.}, $ {\sim}10^{-2}$~m
for the Earth as the central body),
a typical initial radius of the planetary orbit~$ R_0 $
(\textit{e.g.}, $ {\sim}10^9$~m for the Moon moving around the Earth),
and de~Sitter radius~$ r_{\Lambda} $ ($ {\sim}10^{27}$~m, which depends
on the amount of dark energy in the Universe).\footnote{
It was emphasized for the first time by Balaguera-Antol{\'i}nez
\textit{et al.}~\cite{bal06} that
the specific interplay between $ r_g $ and $ r_{\Lambda} $ can result
in a manifestation of the $ \Lambda $-term effects at the spatial scales
much less than~$ r_{\Lambda} $; but that consideration was performed
for the static Kottler metric~(\ref{eq:Kottler_metric}). Besides,
the ``small-scale'' cosmological effects were found also in the collapsing
matter overdensities (\textit{e.g.}, paper~\cite{mot04} and references
therein); but such analyses were performed in the models of ``dynamical''
dark energy and, therefore, irrelevant to the present study.}

To avoid misunderstanding, let us emphasize that the above formulas
were written up to the terms of the first order of~$ r_g $ and
$ r_{\Lambda}^{-1} $. However, we did not assume that the mixed
products~$ r_g \, r_{\Lambda}^{-1} $ are the quantities of
the higher order of smallness and also keep them in the equations,
because the relation between~$ r_g / R_0 $ and $ R_0 / r_{\Lambda} $
can be very different in the various astrophysical situations.
Just this approximation will be used later for the numerical
integration of orbits in Section~2.4.
Besides, our analytical perturbation scheme presented in the next
Section~2.3 also assumes that~$ r_g $ and $ r_{\Lambda}^{-1} $ are
the independent small quantities.

\subsection*{2.3. Perturbative analysis}

The presence of the above-mentioned small ratios $ r_g / R_0 $
and $ R_0 / r_{\Lambda} $ suggests using the perturbation
theory for analyzing the set of
equations~(\ref{eq_motion_1})--(\ref{eq_motion_3}).
However, a choice of the particular perturbation scheme is quite problematic
from the viewpoint of reasonable convergence of the resulting expansions.

We shall restrict our analysis here by the simplest case of
the purely circular initial orbits.
First of all, assuming $ \Lambda \! \to 0 $ (and, consequently,
$ r_{\Lambda} \! \to \infty $), we get the set of equations for
the unperturbed orbit:
\begin{subequations}
\begin{eqnarray}
&& 2 \: \Big( 1 - \frac{r_g}{r} \Big) \: \ddot{t} \:
+ \: 2 \, \frac{r_g}{r^2} \, \dot{t} \, \dot{r} \:
= \: 0 \, ,
\label{eq_unpert_motion_1}
\\[0.5ex]
&& 2 \, \Big( 1 + \frac{r_g}{r} \Big) \, \ddot{r} \:
+ \: \frac{r_g}{r^2} \, {\dot{t}}^{\, 2} \!
- \: \frac{r_g}{r^2} \, {\dot{r}}^{\, 2} \!
- \, 2 \, r {\dot{\varphi}}^{\, 2} = \: 0 \, ,
\label{eq_unpert_motion_2}
\\[0.5ex]
&& r \ddot{\varphi} \: + \, 2 \, \dot{r} \dot{\varphi} \: = \: 0 \, .
\label{eq_unpert_motion_3}
\end{eqnarray}
\end{subequations}

Next, seeking for the solution with constant unperturbed radius $ r \! = r_0 $
(and, consequently, $ \dot r = 0 $, $ \ddot r = 0 $),
we get from equation~(\ref{eq_unpert_motion_1}) that
$ \ddot t = 0 $, \textit{i.e.},
\begin{equation}
t = \, t_0 + \, k \tau \, .
\label{eq_t_perturb}
\end{equation}

Similarly, equation~(\ref{eq_unpert_motion_3}) gives
$ \ddot \varphi = 0 $ and, consequently,
\begin{equation}
\varphi = \, {\varphi}_0 + \, \omega \tau \, .
\label{eq_phi_perturb}
\end{equation}

Without a loss in generality, the initial time~$ t_0 $ and
the initial angle~$ {\varphi}_0 $ can be taken to be zero.
The coefficient~$ k $ represents, evidently, a relativistic correction for
time of the moving particle, and $ \omega $~is the frequency of its revolution.

Substituting~(\ref{eq_t_perturb}) and (\ref{eq_phi_perturb})
into~(\ref{eq_unpert_motion_2}), we get a relation between
the orbital radius and frequency:
\begin{equation}
2 \, r_0^3 \, ( {\omega} / k )^2 = r_g \, ,
\label{eq_3rd_Kepler_law}
\end{equation}
which is an analog of the 3rd Kepler law.
Let us emphasize that $ \omega $~is the frequency of rotation
in terms of the proper time of the moving particle, while
the frequency for an ``external observer'' will be~$ \omega / k \, $.

Taking into account the above-mentioned characteristics of
the unperturbed orbit, we shall seek parameters of the perturbed
orbit in the form:
\begin{subequations}
\begin{eqnarray}
&& r = r_0 ( 1 + \xi ) \, ,
\\
&& t = k \tau ( 1 + \eta ) \, ,
\\
&& \varphi = \omega \tau ( 1 + \zeta ) \, ,
\end{eqnarray}
\end{subequations}
where $ \xi $, $ \eta $, and $ \zeta $ are the functions of~$ \tau $, which
are assumed to be small as compared to unity.
Substituting the above-written expressions into the original set of
equations~(\ref{eq_motion_1})--(\ref{eq_motion_3}) and keeping only
the first-order terms with respect to $ \xi $, $ \dot{\xi} $,
$ \ddot{\xi} $, $ \eta $, $ \dot{\eta} $, $ \ddot{\eta} $,
$ \zeta $, $ \dot{\zeta} $, $ \ddot{\zeta} $, and $ r_{\Lambda}^{-1} $,
we get the following set of equations for the orbital perturbations:
\begin{subequations}
\begin{eqnarray}
&& \big( 1 - r_g r_0^{-1} \big) \, \tau \ddot{\eta} +
2 \big( 1 - r_g r_0^{-1} \big) \, \dot{\eta} +
r_g r_0^{-1} \dot{\xi} +
r_g r_0^{-1} r_{\Lambda}^{-1} k = 0 \, ,
\label{eq_pert_1}
\\
&& 2 \big( 1 + r_g r_0^{-1} \big) \, \ddot{\xi} +
r_g r_0^{-3} k^2 \big[
2 \eta + 2 \tau \dot{\eta} - 2 \zeta - 2 \tau \dot{\zeta} -
3 \xi - 3 r_{\Lambda}^{-1} k \tau \big] = 0 \, ,
\label{eq_pert_2}
\\
&& \tau \ddot{\zeta} + 2 \dot{\zeta} + 2 \dot{\xi} +
2 r_{\Lambda}^{-1} k = 0 \, .
\label{eq_pert_3}
\end{eqnarray}
\end{subequations}

Unfortunately, since the coefficients of this linear system of differential
equations are not constant but depend on~$ \tau $ (which results from
the time-dependent cosmological asymptotics), its analytic treatment
is not an easy task, and no explicit solution can be obtained for the entire
time interval (as distinct from the static Kottler metric).
Therefore, we shall seek for the required perturbations in the form of
a power-series expansion, assuming that $ ( \tau / r_0 ) $ is a small quantity:
\begin{subequations}
\begin{eqnarray}
&& \xi = {\alpha}_1 ( \tau / r_0 )  + {\alpha}_2 ( \tau / r_0 )^2 +
{\alpha}_3 ( \tau / r_0 )^3 + \dots \, ,
\label{eq_xi_expansion}
\\
&& \eta = {\beta}_1 ( \tau / r_0 )  + {\beta}_2 ( \tau / r_0 )^2 +
{\beta}_3 ( \tau / r_0 )^3 + \dots \, ,
\label{eq_eta_expansion}
\\
&& \zeta = {\gamma}_1 ( \tau / r_0 )  + {\gamma}_2 ( \tau / r_0 )^2 +
{\gamma}_3 ( \tau / r_0 )^3 + \dots \, .
\label{eq_zeta_expansion}
\end{eqnarray}
\end{subequations}
We have omitted the terms of zero order because, by definition,
$ \xi (0) = 0 $, $ \eta (0) = 0 $, and $ \zeta (0) = 0 $
(the perturbations are absent at the initial instant of time).
Besides, since~(\ref{eq_pert_1})--(\ref{eq_pert_3}) is
the set of differential equations of the second order,
in general, it is necessary to specify also the first derivatives
$ \dot{\xi} (0) $, $ \dot{\eta} (0) $, and $ \dot{\zeta} (0) $.
However, it is not clear in advance if all such derivatives
are independent of each other, because there is some kind of
degeneracy in the mathematical problem under consideration:
the highest-order derivatives in equations~(\ref{eq_pert_1})
and (\ref{eq_pert_3}) contain the coefficients that vanish
at $ \tau \! = 0 $. Therefore, generally speaking, the number of
initial conditions may be less than three. We shall return later
to the discussion of this subject.

Substituting expansions~(\ref{eq_xi_expansion})--(\ref{eq_zeta_expansion})
into~(\ref{eq_pert_1})--(\ref{eq_pert_3})
and keeping, for example, only the terms up to the second order of
smallness, we get the following set of equations:
\begin{subequations}
\begin{eqnarray}
&& \big[ 1 - ( r_g / r_0 ) \big] \, \Big\{ 2 {\beta}_1 +
6 {\beta}_2 ( \tau / r_0 ) + 12 \, {\beta}_3 ( \tau / r_0 )^2 + \dots \Big\}
\nonumber \\
&& \qquad
+ \, ( r_g / r_0 ) \, \Big\{ {\alpha}_1 + 2 {\alpha}_2 ( \tau / r_0 ) +
3 {\alpha}_3 ( \tau / r_0 )^2 + \dots \Big\}
\nonumber \\
&& \qquad
+ \, ( r_g / r_0 ) ( r_0 / r_{\Lambda} ) k \, = \, 0 \, ,
\\[1ex]
&& \big[ 1 + ( r_g / r_0 ) \big] \, \Big\{ 4 {\alpha}_2 +
12 \, {\alpha}_3 ( \tau / r_0 ) + 24 \, {\alpha}_4 ( \tau / r_0 )^2 + \dots
\Big\}
\nonumber \\
&& \qquad
+ \, ( r_g / r_0 ) k^2 \Big\{ 4 {\beta}_1 ( \tau / r_0 ) +
6 {\beta}_2 ( \tau / r_0 )^2 + \dots
\nonumber \\
&& \qquad
- \, 4 {\gamma}_1 ( \tau / r_0 ) - 6 {\gamma}_2 ( \tau / r_0 )^2 + \dots
\nonumber \\
&& \qquad
- \, 3 {\alpha}_1 ( \tau / r_0 ) - 3 {\alpha}_2 ( \tau / r_0 )^2 + \dots \Big\}
\nonumber \\
&& \qquad
-3 \, ( r_g / r_0 ) ( r_0 / r_{\Lambda} ) k^3 ( \tau / r_0 ) \, = \, 0 \, ,
\\[1ex]
&& {\gamma}_1 + 3 {\gamma}_2 ( \tau / r_0 ) + 6 {\gamma}_3 ( \tau / r_0 )^2 +
\dots
\nonumber \\
&& \qquad
+ \, {\alpha}_1 + 2 {\alpha}_2 ( \tau / r_0 ) + 3 {\alpha}_3 ( \tau / r_0 )^2 +
\dots
\nonumber \\
&& \qquad
+ \, ( r_0 / r_{\Lambda} ) k \, = \, 0 \, .
\end{eqnarray}
\end{subequations}

Next, equating the terms with the same powers of~$ ( \tau / r_0 ) $ to zero,
we get an infinite set of linear algebraic equations for the determination of
coefficients of expansions~(\ref{eq_xi_expansion})--(\ref{eq_zeta_expansion}):
\begin{subequations}
\begin{eqnarray}
&& 2 \, \big[ 1 - ( r_g / r_0 ) \big] \, {\beta}_1 + ( r_g / r_0 ) {\alpha}_1 +
( r_g / r_0 ) ( r_0 / r_{\Lambda} ) k = 0 \, , \:
\\[-0.5ex]
&& {\alpha}_2 = 0 \, ,
\\[-0.5ex]
&& {\gamma}_1 + {\alpha}_1 + ( r_0 / r_{\Lambda} ) k = 0 \, ;
\end{eqnarray}
\end{subequations}
\vspace*{-5ex}
\begin{subequations}
\begin{eqnarray}
&& 6 \, \big[ 1 - ( r_g / r_0 ) \big] \, {\beta}_2 +
2 ( r_g / r_0 ) {\alpha}_2 = 0 \, ,
\\[-0.5ex]
&& 12 \, \big[ 1 + ( r_g / r_0 ) \big] \, {\alpha}_3 +
4 ( r_g / r_0 ) k^2 {\beta}_1 - \, 4 ( r_g / r_0 ) k^2 {\gamma}_1
\nonumber \\[-0.5ex]
&& \qquad
- \, 3 ( r_g / r_0 ) k^2 {\alpha}_1
- 3 ( r_g / r_0 ) ( r_0 / r_{\Lambda} ) k^3 = 0 \, ,
\\[-0.5ex]
&& 3 {\gamma}_2 + 2 {\alpha}_2 = 0 \, ;
\qquad \qquad \qquad \qquad \qquad \qquad \qquad \qquad \quad
\end{eqnarray}
\end{subequations}
\vspace*{-5ex}
\begin{subequations}
\begin{eqnarray}
&& 12 \, \big[ 1 - ( r_g / r_0 ) \big] \, {\beta}_3 +
3 ( r_g / r_0 ) {\alpha}_3 = 0 \, ,
\\[-0.5ex]
&& 24 \, \big[ 1 + ( r_g / r_0 ) \big] \, {\alpha}_4 +
6 ( r_g / r_0 ) k^2 {\beta}_2
\nonumber \\[-0.5ex]
&& \qquad
- \, 6 ( r_g / r_0 ) k^2 {\gamma}_2 - 3 ( r_g / r_0 ) k^2 {\alpha}_2 = 0 \, ,
\\[-0.5ex]
&& 6 {\gamma}_3 + 3 {\alpha}_3 = 0 \, ;
\\[2ex]
&& etc.
\qquad \qquad \qquad \qquad \qquad \qquad \qquad \qquad \qquad \qquad \qquad
\nonumber
\end{eqnarray}
\end{subequations}

As follows from the analysis of this system, it can be uniquely solved
if one of the expansion parameters (for example, $ {\alpha}_1 $)
is specified in advance.
Then, all other parameters will be determined by the recursion
relations:
\begin{subequations}
\begin{eqnarray}
&& {\beta}_1 = - (1/2 ) \, \big[ 1 - ( r_g / r_0 ) \big]^{-1} ( r_g / r_0 )
\big[ {\alpha}_1 + ( r_0 / r_{\Lambda} ) k \, \big] \, ,
\\[-0.2ex]
&& {\gamma}_1 = - {\alpha}_1 - ( r_0 / r_{\Lambda} ) k \, ,
\\[-0.2ex]
&& {\alpha}_2 = 0 \, , \quad {\beta}_2 = 0 \, , \quad {\gamma}_2 = 0 \, ,
\\[-0.2ex]
&& {\alpha}_3 = (1/12) \, \big[ 1 + ( r_g / r_0 ) \big]^{-1} ( r_g / r_0 ) k^2
\nonumber \\[-0.2ex]
&& \qquad \; \times
\big[ 3 ( r_0 / r_{\Lambda} ) k + 3 {\alpha}_1 - 4 {\beta}_1 + 4 {\gamma}_1
\big] \, ,
\\[-0.2ex]
&& {\beta}_3 = - (1/4) \, \big[ 1 - ( r_g / r_0 ) \big]^{-1}
( r_g / r_0 ) {\alpha}_3 \, ,
\\[-0.2ex]
&& {\gamma}_3 = - (1/2) {\alpha}_3 \, ,
\\[-0.2ex]
&& etc.
\nonumber
\end{eqnarray}
\end{subequations}
Therefore, as was already mentioned in the previous discussion,
only one initial condition for the first derivatives should be
specified, \textit{e.g.}, $ \dot{\xi} (0) = {\alpha}_1 / r_0 $.
Unfortunately, this condition cannot be prescribed from some
general principles, and its specification requires consideration
of the additional physical processes, \textit{e.g.}, the details of
formation and evolution of the planetary system.

On the other hand, if we do not wish to go into the above-mentioned
details, a reasonable formulation of the problem may be
as follows. Let us arbitrarily take $ \dot{\xi} (0) = 0 $
(\textit{i.e.}, $ {\alpha}_1 = 0 $).
Since $ \xi $~is a perturbation of the Robertson--Walker
radial distance~$ r $ (which is co-moving with a cosmological
background), the zero value of its derivative means that
initially the test particle is assumed to be fully dragged
by the cosmological expansion. Next, it is interesting to pose
the question:
\textit{Will this particle be completely detached from
the cosmological background if the local gravitational field
by the central body is sufficiently strong?}
(As was already mentioned in the end of Section~2.2.,
in the realistic planetary systems, the characteristic magnitude of
the local gravitational force is very large in comparison
with the magnitude of cosmological influences associated with
$ \Lambda $-term: $ ( r_g / r_0 ) \gg ( r_0 / r_{\Lambda} ) $.)
If such suppression of the cosmological expansion by the local
gravity really takes place, then the perturbation~$ \xi $
should experience a secular decrease with time (\textit{i.e.},
possess a negative average derivative), and the rate of such
decrease should be sufficient to compensate the rate of
cosmological expansion after a transformation from
the Robertson--Walker to the measurable physical distance.
Otherwise, if $ \xi $ does not experience the secular decrease
or possesses the decrease with insufficient rate, we should
conclude that the cosmological expansion survives even under
the action of very strong local gravity.

To answer the above-posed question, it is necessary to study
behavior of the radial perturbation~$ \xi $ during a large
number of revolutions. Unfortunately, the power
series~(\ref{eq_xi_expansion})--(\ref{eq_zeta_expansion})
used in our analysis are poorly suited for this aim
because of the limited range of convergence: taking only
a few first terms of the expansion enables us to describe only
a fraction of a single revolution. Therefore, we need to employ
a numerical integration of the equations of motion.

\subsection*{2.4. Numerical integration}

Since the analytical treatment performed in the previous section is
limited to the case of quite small time intervals (and also the purely
circular initial orbits), we shall use the numerical procedures
to overcome these restrictions.

First of all, it is convenient to introduce the formal six-dimensional
vector
\begin{equation}
{\bf x} = \big( t, \dot{t}, r, \dot{r}, {\varphi}, \dot{\varphi} \big)
\label{eq_def_x}
\end{equation}
and to rewrite the original set of
equations~(\ref{eq_motion_1})--(\ref{eq_motion_3}) as a system of
six first-order differential equations resolved with respect to
the derivatives. Keeping only the terms of the same order of smallness
as before, we get:
\begin{subequations}
\begin{eqnarray}
\dot{x_1} & = & x_2 \, ,
\label{eq_dif_x1}
\\
\dot{x_2} & = & - \frac{1}{2} \, \frac{r_g}{r_{\Lambda}} \, \frac{x_2^2}{x_3}
  - \frac{r_g}{x_3^2} \Big( 1 - \frac{x_1}{r_{\Lambda}} \Big) x_2 \, x_4
\nonumber \\ &&
  - \frac{1}{r_{\Lambda}} \Big( 1 + \frac{3}{2} \, \frac{r_g}{x_3} \Big) x_4^2
  - \frac{x_3^2}{r_{\Lambda}} \Big( 1 - \frac{r_g}{x_3} \Big) x_6^2 \, ,
\label{eq_dif_x2}
\\
\dot{x_3} & = & x_4 \, ,
\label{eq_dif_x3}
\\
\dot{x_4} & = & - \frac{1}{2} \, \frac{r_g}{x_3^2}
    \Big( 1 - 3 \frac{x_1}{r_{\Lambda}} \Big) x_2^2
  - \frac{2}{r_{\Lambda}} \Big( 1 - \frac{1}{2} \, \frac{r_g}{x_3} \Big)
    x_2 \, x_4
\nonumber \\ &&
  + \frac{1}{2} \, \frac{r_g}{x_3^2} \Big( 1 - \frac{x_1}{r_{\Lambda}} \Big)
    x_4^2
  + \Big[ 1 - \frac{r_g}{x_3} \Big( 1 + 3 \frac{x_1}{r_{\Lambda}} \Big) \Big]
    x_3 \, x_6^2 \, ,
\label{eq_dif_x4}
\\
\dot{x_5} & = & x_6 \, ,
\label{eq_dif_x5}
\\
\dot{x_6} & = & - \frac{2}{r_{\Lambda}} \, x_2 \, x_6
  - \frac{2}{x_3} \, x_4 \, x_6 \, ,
\label{eq_dif_x6}
\end{eqnarray}
\end{subequations}

From the formal point of view, the above set of equations is suitable
for integration by any standard software. However, in practice,
such integration turns out to be challenging, because the standard
accuracy of representation of the floating-point numbers in a computer
is insufficient to cover the realistic range of parameters, discussed
in the end of Section~2.2. As a result, it is
necessary to use a special software for emulation of the high-accuracy
arithmetic. A detailed numerical study of the
equations~(\ref{eq_dif_x1})--(\ref{eq_dif_x6})
will be published elsewhere; while here we restrict our consideration by
a few toy models, where difference between the characteristic scales of
the problem is not so much as in reality. This will help us to reveal
the most important qualitative features of the resulting motion.

From here on, it is convenient to take the typical initial size of
the orbit (\textit{e.g.}, its major semi-axis)~$ R_0 $ as the unit of length.
Then, the unit of time will be $ R_0 / c $ (or just $ R_0 $
if $ c \equiv 1 $, as in the previous equations).
The corresponding dimensionless variables,
normalized by~$ R_0 $ and~$ R_0 / c $, will be denoted by asterisks.
Besides, it should be kept in mind that all the quantities used here
refer to the Robertson--Walker coordinate system (\textit{i.e.},
co-moving with the cosmological background).
Therefore, to answer the question of significance of the cosmological
influences, we should return back to the measurable coordinates,
denoted by primes.

%%%%%%%%%%%%%%%%%%%%%%%%%%%%%%%%%%%%%
\begin{figure}[t]
\begin{center}
\includegraphics[width=12cm]{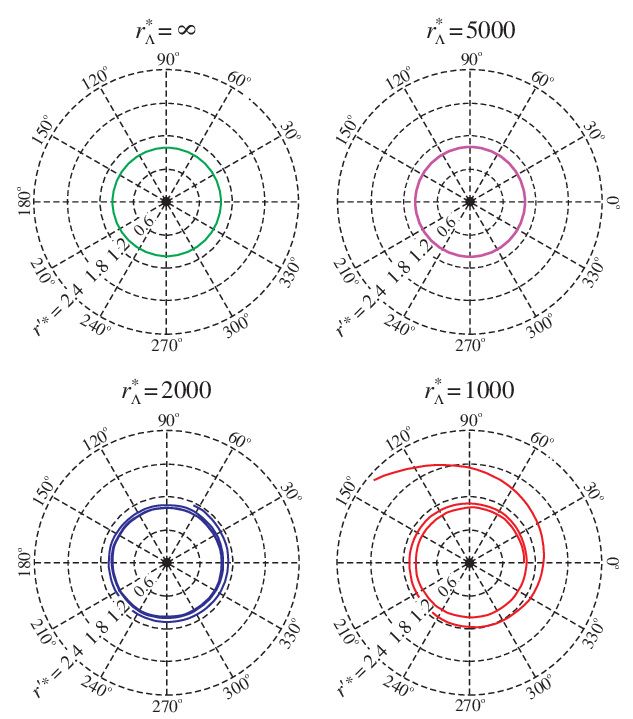}
\end{center}
\caption{\label{fig:orbits}
Orbits of a test particle at the specified Schwarzschild
radius of the central body~$ r_g^* = 0.01 $ and various
de~Sitter radii~$ r_{\Lambda}^* \, $
(\textit{i.e.}, the various values of the $ \Lambda $-term).}
\end{figure}
%%%%%%%%%%%%%%%%%%%%%%%%%%%%%%%%%%%%%

Let us take, for example, $ r_g^* = 0.01 $ and study the characteristic
shapes of the test-particle orbits at various values of~$ r_{\Lambda}^* $.
The results of numerical integration of the
equations~(\ref{eq_dif_x1})--(\ref{eq_dif_x6}) for a few values
of de~Sitter radius are shown in Fig.~\ref{fig:orbits}.
If $ r_{\Lambda}^* = \infty $ (\textit{i.e.}, $ \Lambda = 0 $),
the orbit is closed, as should be evidently expected.
Next, when the values of~$ r_{\Lambda}^* $ decrease down to
a few thousand (\textit{i.e.}, the values of~$ \Lambda $ increase),
the orbits become slightly spiral; and such unwinding is
expressed very well, for example, at $ r_{\Lambda}^* = 1000 $.
Let us emphasize that a relative magnitude of the cosmological
influences in this case still remains much less than the relative
magnitude of the local gravitational forces:
$ 1 / r_{\Lambda}^* = ( R_0 / r_{\Lambda} ) = 0.001 \ll
r_g^* = ( r_g / R_0 ) = 0.01 $.
Nevertheless, \textit{
the effect of cosmological perturbations is accumulated
and becomes significant in the course of a few revolutions}.

%%%%%%%%%%%%%%%%%%%%%%%%%%%%%%%%%%%%%
\begin{figure}[t]
\begin{center}
\includegraphics[width=8.5cm]{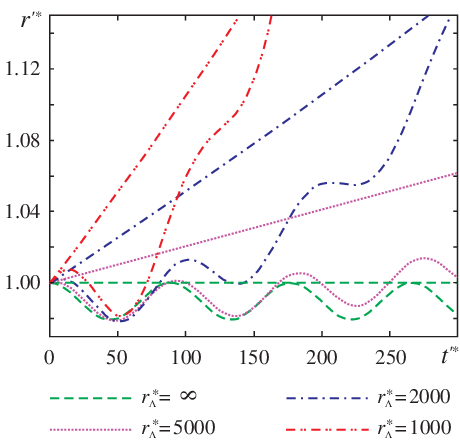}
\end{center}
\caption{\label{fig:r_t}
Radii of the orbits as functions of time at the specified
Schwarzschild radius of the central body $ r_g^* = 0.01 $
and various de~Sitter radii~$ r_{\Lambda}^* $ (wavy curves)
vs.\ the radii of the test particle in the standard Hubble
flow (straight lines).}
\end{figure}
%%%%%%%%%%%%%%%%%%%%%%%%%%%%%%%%%%%%%

The same phenomenon is presented in a different way in
Fig.~\ref{fig:r_t}, where the orbital radius~$ {r'}^* $ is plotted
as function of the coordinate time~$ {t'}^* $ (which differs,
in fact, only slightly from the proper time of the moving particle).
The curves are wavy just because the initial unperturbed orbit
was taken to be slightly elliptic.
The straight lines in this figure represent the standard Hubble
motion, which would be experienced by the test particle without
the central gravitating body at the same values of~$ r_{\Lambda}^* $.
(We should keep in mind that the smaller values of de~Sitter radius
correspond to the larger values of $ \Lambda $-term,
because $ r_{\Lambda} = \sqrt{ 3 / \Lambda } $.)

Let us pay attention, particularly, to the case $ r_{\Lambda}^* = 2000 $,
which is presented by the dash-and-dotted curve.
As is seen, the local gravitational force by the central body
initially suppresses the cosmological expansion of the orbit,
but after a number of revolutions the cosmological influence is
accumulated, and the resulting rate of secular increase in
the orbital radius tends to the rate of the standard Hubble flow.

\section*{3. Discussion and conclusions}

Let us recall the main question, posed in the very end of Section~2.3:
Can a sufficiently strong local gravitational force (produced by
the central mass~$ M $ and characterized by its Schwarzschild
radius~$ r_g $) completely suppress the effect of expanding
cosmological background (characterized, in the simplest model,
by the inverse de~Sitter radius~$ 1 / r_{\Lambda} $)?
The most nontrivial conclusions, following from the numerical
treatment of a few toy cases in Section~2.4, are:
(i)~if the local gravitational attraction is sufficiently strong,
it does initially suppress the cosmological recession of the test
particle;
(ii)~however, the cosmological influences are accumulated during
a number of revolutions, and the recession rate is gradually
restored up to the value comparable to the standard Hubble
flow at infinity. In our opinion, this points to
the potential importance of cosmological effects in the dynamics
of small-scale (\textit{e.g.}, planetary) gravitationally-bound
systems (at least, in the case when the cosmological background
is formed by the perfectly-uniform $ \Lambda $-term).

Of course, a much more careful analysis should be performed to draw
the ultimate conclusion on this subject.
First of all, it is necessary to include into the equations
of motion a lot of additional physical factors affecting
the planetary dynamics (such as the mutual attraction between the planets,
the tidal force between them, \textit{etc.}).
Besides, one should find a way to specify unambiguously the initial
conditions for these equations (which were taken somewhat arbitrarily in
the present paper). Probably, this will require to consider the entire
process of formation of the planetary systems.

It is interesting to mention that there is some empirical evidence in
favor of the cosmological expansion in the Earth--Moon system:
This is a well-known disagreement between
the rates of secular increase in the lunar semi-major axis measured,
on the one hand, immediately by the lunar laser ranging~\cite{nor99},
$ \dot{R}_{\rm LLR} = 3.8 \pm 0.1 $~cm/yr~\cite{dic94}, and,
on the other hand, derived indirectly from the data on
the Earth's rotation deceleration,
$ \dot{R}_{\rm rot} = 1.6 \pm 0.2 $~cm/yr
(see, for example, the time series compiled in the monograph~\cite{sid02}).
Surprisingly, these two values can be reconciled with each other
quite well if, along with the commonly-considered tidal
interaction between the Earth and Moon, one takes into account
also a contribution from the local Hubble expansion; and the rate
of such an expansion turns out to be in reasonable agreement
with the large-scale cosmological data~\cite{dum08}.

However, it is commonly believed that the modern solar-system
ephemerides are able to explain all the planetary motions
without any additional cosmological influences, apart from
the well-known post-Newtonian corrections (\textit{e.g.},
papers~\cite{pit05a,pit05b} and references therein). If this is
really the case, and no further corrections for the $ \Lambda $-term
are necessary, then the high-accuracy planetary observations might
be used to impose strong constraints on the amount of dark energy
represented by the perfectly uniform $ \Lambda $-term.
So, the models with the ``dynamic'' dark energy, described by
a scalar field and/or the nontrivial equation of state, may become
preferable.

In summary, we presented a rigorous mathematical formulation of
the two-body problem in the $ \Lambda $-dominated Universe with
the adequate cosmological asymptotics at infinity. A set of solutions
of the respective equations of motion was obtained numerically.
It was found that, at least in some circumstances, the specific secular
perturbations of orbits by the $ \Lambda $-term become appreciable
and can even reach the rate of the standard Hubble flow.
From our point of view, this fact is interesting both by itself
and may have important consequences for the long-term dynamics of
the realistic Keplerian orbits. Therefore, a possible presence of
such effects should be taken into account very carefully in the future
high-precision analyses of planetary and interacting stellar systems.

\acknowledgments

I am grateful to
Yu.~Baryshev,
N.~Capitaine,
M.L.~Fil'chenkov,
S.S.~Gerstein,
C.~L{\"a}m\-merzahl,
S.A.~Klioner,
S.M.~Kopeikin,
J.~M{\"u}ller,
K.~Nordtvedt,
M.~Nowakowski,
E.V.~Pitjeva, and
A.V.~Toporensky
for valuable discussions and critical comments.
I am also grateful to Wilhelm und Else Heraeus-Stiftung
for the opportunity to present this work at a few workshops.

\end{document}